\documentclass[%
reprint,
superscriptaddress,
 amsmath,amssymb,
 aps,
floatfix,
]{revtex4-1}

\usepackage{graphicx}
\usepackage{dcolumn}
\usepackage{bm}
\usepackage{siunitx}
\usepackage{gensymb}

\usepackage{amsmath} 

\usepackage[caption=false]{subfig}
\usepackage{natbib}
\usepackage{xcolor}
\usepackage{hyperref}

\begin{document}


\title{Distortions to the penetration depth and coherence length of superconductor/normal-metal superlattices}%
 
\author{P. Quarterman}
\email{patrick.quarterman@nist.gov}

\affiliation{NIST Center for Neutron Research, National Institute of Standards and Technology, Gaithersburg, MD 20899, USA}

\author{Nathan Satchell}
\email{N.D.Satchell@leeds.ac.uk}

\affiliation{Department of Physics and Astronomy, Michigan State University, East Lansing, Michigan 48912, USA}
\affiliation{School of Physics and Astronomy, University of Leeds, Leeds LS2 9JT, United Kingdom}

\author{B. J. Kirby}
\affiliation{NIST Center for Neutron Research, National Institute of Standards and Technology, Gaithersburg, MD 20899, USA}

\author{Reza~Loloee}
\affiliation{Department of Physics and Astronomy, Michigan State University, East Lansing, Michigan 48912, USA}

\author{Gavin Burnell}
\affiliation{School of Physics and Astronomy, University of Leeds, Leeds LS2 9JT, United Kingdom}

\author{Norman O. Birge}
\affiliation{Department of Physics and Astronomy, Michigan State University, East Lansing, Michigan 48912, USA}

\author{J. A. Borchers}
\affiliation{NIST Center for Neutron Research, National Institute of Standards and Technology, Gaithersburg, MD 20899, USA}

\date{\today}

\begin{abstract}
  Superconducting (\textit{S}) thin film superlattices composed of Nb and a normal metal spacer (\textit{N}) have been extensively utilized in Josephson junctions given their favorable surface roughness compared to Nb films of comparable thickness. In this work, we characterize the London penetration depth and Ginzburg-Landau coherence lengths of \textit{S}/\textit{N} superlattices using polarized neutron reflectometry and electrical transport. Despite the normal metal spacer layers being only approximately 8\% of the total superlattice thickness, we surprisingly find that the introduction of these thin \textit{N} spacers between \textit{S} layers leads to a dramatic increase in the measured London penetration depth compared to that of a single Nb film of comparable thickness.
 Using the measured values for the effective in- and out-of-plane coherence lengths, we quantify the induced anisotropy of the superlattice samples and compare to a single Nb film sample. From these results, we find that that the superlattices behave similarly to layered 2D superconductors.
\end{abstract}
\maketitle


\section{Introduction}

Superconducting materials have long been of interest since they were first discovered in 1911 by Kammerlingh Onnes \cite{Onnes}. The two fundamental properties most closely associated with superconductivity are zero electrical resistance and the expulsion of magnetic field, where the latter is known as the Meissner effect \cite{Meissner}. The two typical lengthscales describing superconductivity are the coherence length ($\xi$) and the penetration depth ($\lambda$), which we set out to measure directly in this work. The ratio of these two length scales, known as the Ginzburg-Landau parameter ($\kappa=\lambda/\xi$), governs whether the superconductor is type-I or type-II, where a type-I is defined as $\kappa < 1 / \sqrt{2}$ and type-II when $\kappa > 1 / \sqrt{2}$. In bulk single crystal, Nb is a borderline type-I/type-II superconductor with $\xi \approx  \lambda \approx 41$ nm \cite{kittel1976introduction}. Upon reducing dimensionality and introducing the disorder associated with polycrystalline thin film growth, Nb becomes strongly type-II with a typical $\kappa$ between 8 and 10.  

Superconducting technologies based on Josephson junctions are a promising candidate for a low power computational alternative to traditional CMOS technologies \cite{6449287,soloviev2017beyond}. Of particular interest, here, are ferromagnetic Josephson junctions which form the memory bits in such a scheme \cite{bell_controllable_2004, 1439786, baek2014hybrid,  doi:10.1063/1.4862195, PhysRevApplied.3.011001,  gingrich_controllable_2016, dayton2017experimental, PhysRevB.97.024517, Glickeaat9457,  Madden_2018, 10.1117/12.2321109,8911211, doi:10.1063/1.5140095}. There have been numerous reports concerning the supercurrent passing through ferromagnetic Josephson junctions, including the discovery of spin triplet pair correlations in these systems \cite{PhysRevLett.86.4096,keizer2006spin,PhysRevLett.104.137002,robinson2010controlled,PhysRevB.82.100501,PhysRevLett.116.077001,lahabi2017controlling,doi:10.1098/rsta.2015.0150}. In order to improve the properties of the thin ferromagnetic layers in the Josephson junctions, it is important that the surface roughness of the Nb electrode be as small as possible \cite{8115303}. It has long been known that introducing a thin Al layer between Nb films improves the surface roughness compared to Nb films of equivalent thickness; this is understood to be due to the Al forming amorphously and thus preventing columnar growth of Nb \cite{ NbSiplanarization,NbAlbilayer,NbAlplanarization}. Given that the Al is sufficiently thin in these multilayers, the Al layer superconducts via proximity effect from the neighboring Nb layers. It has previously been shown that by replacing Nb with a Nb/Al superlattice, the curious effect of non-linear scaling of critical currents with area can be resolved \cite{NbAlarea}. Recently, it was discovered that substituting Al with Au has the same effect on the surface roughness of the superlattice, as we report in this work.

The London penetration depth ($\lambda_\text{L}$) characterizes the depth of penetration of an externally applied field and is the length scale associated to the Meissner effect, hence it is often referred to as the magnetic screening length \cite{kittel1976introduction}. $\lambda_\text{L}$ is typically determined from measurements of flux expulsion by muon implantation \cite{blundell_spin-polarized_1999} or surface microwave \cite{Anlage2007} techniques.  

In superconducting thin film samples, when an \textit{in situ} magnetic field is applied and the samples are cooled below the superconducting transition temperature, polarized neutron reflectometry (PNR) is sensitive to the absence of magnetic field due to the Meissner effect as a function of depth \cite{PenetrationPNR,PenetrationNIST,PNRPbVortex}. The availability of PNR makes this technique highly attractive for the study of thin films with buried interfaces. PNR directly probes the nuclear composition and magnetization, as a function of depth, in thin film systems. The non-spin flip reflectivities ($R^{\uparrow \uparrow}$ and $R^{\downarrow \downarrow}$) are sensitive to the nuclear composition and in-plane magnetization, aligned with the \textit{in situ} magnetic field, as a function of depth through the sample. Han \textit{et al} examined Nb/Al superlattices near and above the lower critical field by measuring the penetration depth ($\lambda_\text{L}$) and characterizing the formation of superconducting vortices using PNR \cite{PNRNbAl}. After the early investigations of field expulsion in superconducting thin film systems, PNR characterization of superconductors became less common due to the unavailability of sufficiently smooth interfaces in thin film materials with novel properties. Early attempts to study thin films of YBCO, for example, had limited success due to the large surface roughness \cite{PenetrationNIST}. Recently, improvements in thin film growth combined with experimental observations and theoretical predictions of proximity effects in superconductor-ferromagnet (\textit{S}/\textit{F}) based heterostructures have renewed interest for characterizing the field expulsion profile in detail \cite{PhysRevX.5.041021,flokstra2016remotely,AnomalousMeissner,PNRNbGd,PhysRevB.100.020505,doi:10.1063/1.5114689,ElectromagneticProximity}. The PNR measurements and analysis performed here are expandable to a wide range of conventional and unconventional superconducting systems, and layered heterostructures containing both superconducting and non-superconducting layers. However, we wish to note that these measurements remain challenging due to the magnetic scattering length density resulting from expelled field of a superconductor being as much as two orders of magnitude lower than that of strong ferromagnets, such as Fe. 

The Ginzburg-Landau coherence length ($\xi_\text{GL}$) characterizes the distance over which superconductivity can vary without undue energy increase. While the Ginzburg-Landau theory is strictly valid only near the critical temperature  ($T_{\text{critical}}$), it has been utilized extensively to describe data over a much broader temperature range \cite{tinkham_introduction_2004}. 
Experimentally, one can estimate $\xi_\text{GL}$ from temperature-dependent electrical transport measurements of the upper critical field ($H_{c2}$). In conventional bulk superconductors, $H_{c2}$ is isotropic. In thin film and layered superconductors, the upper critical field for field orientations parallel ($H_{{c2} \parallel}$) and perpendicular ($H_{{c2} \perp}$) to the sample plane can differ significantly. To determine an accurate estimate for $\xi_\text{GL}$, $H_{c2}$ data must be analyzed in the appropriate geometric limit \cite{tinkham_introduction_2004}.

In this work, we directly determine the effects of introducing a thin (with respect to $\xi_{\text{GL}}$) non-superconducting layer (\textit{N}, which is Al or Au) in Nb-dominant superlattices. Quantitative understanding of $\lambda_\text{L}$ and $\xi_\text{GL}$ are important for modelling and interpreting the behavior of superconducting devices such as Josephson junctions and Superconducting QUantum Interference Devices (SQUIDs), in which the use of superlattices to reduce surface roughness may be advantageous.  For example, in a Josephson junction the characteristic Fraunhofer pattern is determined by the flux ($\Phi$) in the junction, $\Phi =  \mu_{0}H w (2\lambda_\text{L} + d)$, where $w$ is the width of the patterned junction and $d$ is the thickness of the junction. However, this description is modified if $\lambda_\text{L}$ becomes comparable to or longer than the thickness of the superconducting electrode, which is common in thin films \cite{doi:10.1002/352760278X}. As such, an independent and direct measurement of $\lambda_\text{L}$ is important for thorough characterization and understanding of these Josephson junctions. In trying to analyze the inductance of their SQUID devices built upon Nb/Al multilayers, Madden \textit{et al.} found that they could only adequately model their devices by allowing the penetration depth to be about 185 nm, rather than the 85~nm typical value associated with sputtered Nb \cite{Madden_2018}. With PNR, we are able to directly examine the field expulsion to thoroughly evaluate changes in the penetration depth in \textit{S/N} superlattice structures.
We find distinctly different superconducting properties of \textit{S}/\textit{N} superlattice samples compared to a single Nb thin film using PNR and electrical transport measurements. We observe large, consistent modification of $\lambda_\text{L}$ and $\xi_\text{GL}$, which allows us to directly probe and quantify the weakening of the superconducting coupling in the out-of-plane direction associated with the introduction of the thin normal metal intermediary layers.  

We also report that while our Nb film is best characterized as a 3D superconductor, the superlattices are best described at low temperatures by the 2D limit of layered superconductors. Such 2D states have been observed in a number of different \textit{S}/\textit{N} superlattices, but as far as we know not in the composition of the two superlattices studied here \cite{doi:10.1080/00018738900101112}. When a superlattice is made by layering a superconductor with normal metals \cite{NbCuCrossover,V-AgCrossover}, semiconductors \cite{NbGeCrossover,Pb-GeCrossover}, or ferromagnets \cite{NbCo2D}, measurements of $H_{c2}$ (T) have long been known to show a transition from 3D to 2D superconductor behavior when the temperature is reduced \cite{doi:10.1080/00018738900101112}. This crossover 
is understood to occur due to reduction of the effective perpendicular coherence length ($\xi_\perp$), which results in a decoupling of the layers in the superlattice \cite{tinkham_introduction_2004,PhysRevLett.66.2826}. This is a surprising result given that our \textit{N} layers are thin enough that we would expect strong and fully proximitized superconducting coupling across the entirety of the superlattices.

\section{Experimental methods}
 
Thin films are deposited by sputtering with base pressure of $2 \times 10^{-8}$~Torr and partial water pressure of $3 \times 10^{-9}$~Torr ($4 \times 10^{-7}$~Pa), after liquid nitrogen cooling. We grow the films on 12.5 mm x 12.5 mm Si substrates, which have a typical native oxide layer. Growth is performed at an approximate Ar (6N purity) pressure of 2~mTorr and temperature of -25 $^{\degree}$C. Triode sputtering is used for Nb and Al from 57~mm diameter targets, and dc magnetron sputtering is used for Au from a 24~mm diameter target. Materials are deposited at typical growth rates of 0.4~nm~s$^{-1}$ for Nb and Au, and 0.2~nm~s$^{-1}$ for Al from targets with 4N purity. Growth rates are calibrated using an \textit{in situ} quartz crystal film thickness monitor and checked by fitting to Kiessig fringes obtained from X-ray reflectometry (XRR) on reference samples. 

The superlattices had a full stack structure of [\textit{S}(25)/\textit{N}(2.4)]$_{\times 7}$/\textit{S}(25), where the nominal thicknesses are denoted in nanometers, \textit{S} refers to superconducting Nb, and \textit{N} refers to Al or Au. Hereafter, the superlattice samples are referred to as Nb/Al or Nb/Au. The thicknesses of the \textit{S} and \textit{N} layers are guided by previous work \cite{NbAlarea}, and the number of repeats chosen so the films were thick enough to give appreciable magnetic field screening to observe with PNR. In addition, a 200 nm film of Nb was grown for comparison to the superlattices in our current investigation and is comparable to Nb films described in previous studies \cite{PenetrationNIST,PenetrationPNR}. 

We collect PNR using the Polarized Beam Reflectometer and Multi-Angle Grazing-Incidence K-vector reflectometer at the NIST Center for Neutron Research (NCNR). The incident neutron spins are polarized parallel or antiparallel to the applied in-plane magnetic field (\textit{H}) with supermirrors, and reflectivity is measured in the non-spin-flip cross sections ($R^{\uparrow \uparrow}$ and $R^{\downarrow \downarrow}$) as a function of the momentum transfer ($Q$) normal to the film surface. Given the beam incident beam is in the grazing configuration for the entire $Q$ range measured, the neutron beam effectively scatters across the entire sample. The PNR data are reduced and modeled using the REDUCTUS \cite{reductus} software package and model-fit using the REFL1D program \cite{refl1d,refl1dweb}. The uncertainty of each fitting parameter is estimated using a Markov-chain Monte-Carlo simulation implemented by the DREAM algorithm in the BUMPS Python package, and with the number of steps taken allowing for the uncertainty to be reported with two significant digits \cite{DREAM}. Data are gathered at temperatures as low as 3 K, using a closed cycle refrigerator, and with an \textit{in situ} magnetic field of 42 mT applied in the sample plane. To avoid concerns of flux trapping in the sample, we do not change field when below the transition temperature of Nb ($\approx$ 9 K). When changing field states, the temperature is increased to approximately 12 K. For reproducibility of the magnetic field condition, a saturating field of 700 mT is then applied, followed by lowering to the desired measurement field, and finally the sample is cooled to the base temperature of 3 K. XRR and rocking curves are used to confirm the structural model determined from PNR. (An in-depth discussion of the XRR results can be found in the supplemental materials \cite{SM}). 

Electrical transport measurements are performed using a conventional four-point-probe measurement configuration with lock-in amplifier and 100 $\mu$A current source on a cut of the samples with area of approximately 3 mm $\times$ 1 mm, and so we can estimate a current density of $5 \times 10^5 \text{A/m}^2$. We collect transport data in a \textsuperscript{4}He cryostat with variable temperature insert (1.3 - 300~K) and 8~T superconducting solenoid. Our resistance measurements are performed at a fixed temperature by continuously ramping the magnetic field. Resistance as a function of the in- and out-of-plane field, at various temperatures, determines the upper critical fields ($H_{c2\parallel}$, $H_{c2\perp}$). The Ginzburg-Landau coherence lengths in the plane and perpendicular to the plane are extracted by fitting to measurements of $H_{c2}$ as a function of temperature \cite{tinkham_introduction_2004}.

\section{Results}

\subsection{Polarized Neutron Reflectometry}

\begin{figure*}
        \includegraphics[width=0.9\textwidth]{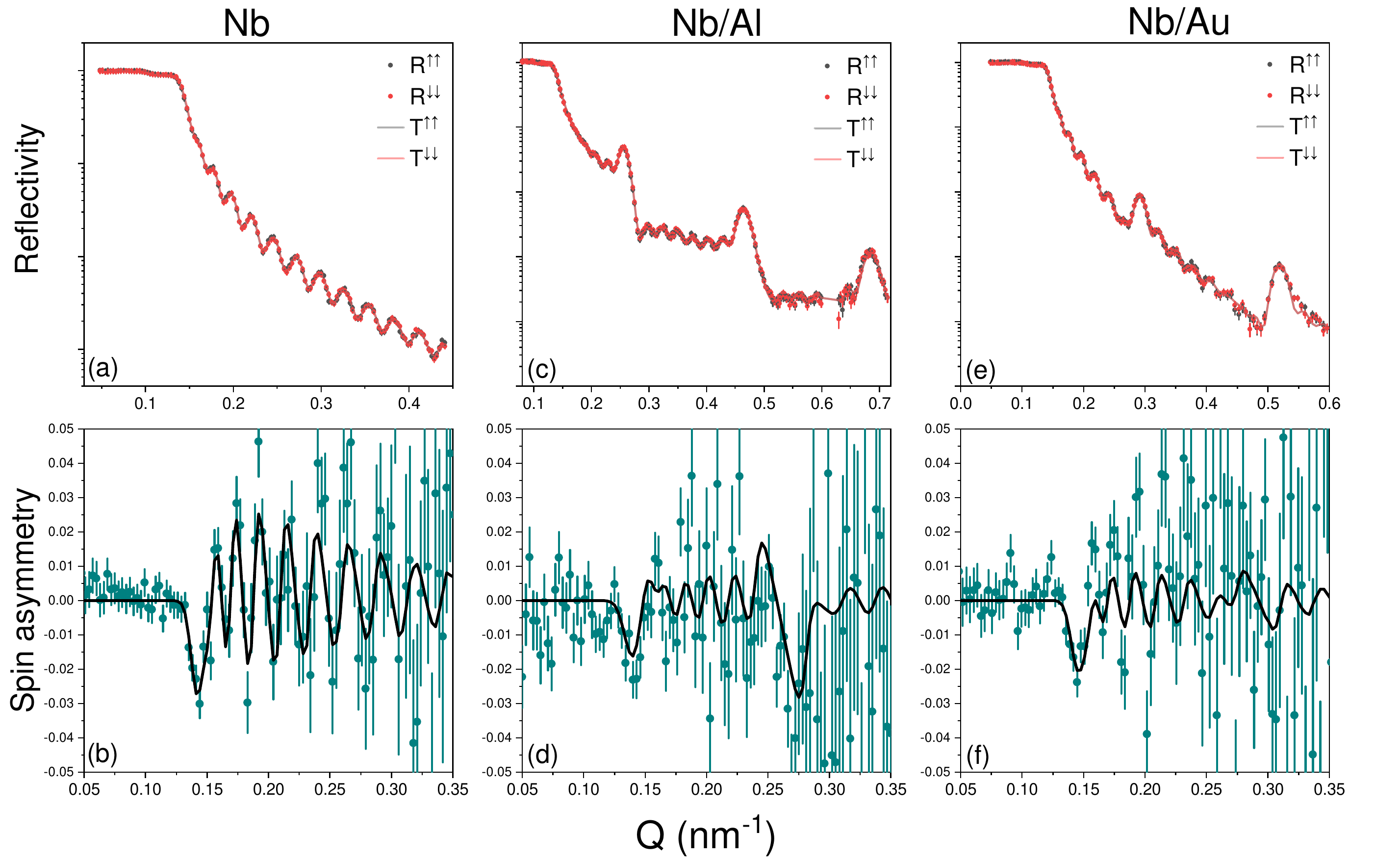}

        \caption{Non-spin flip cross-section PNR data (points) with theoretical fits (T, line) and associated spin asymmetry for (a,b) Nb, (c,d) Nb/Al, and (e,f) Nb/Au superlattices in a field of 42 mT and temperature of 3 K. Error bars are representative of 1 $\sigma$ for the data.}
        \label{fig:PNR}
\end{figure*}

The non-spin flip PNR and the spin asymmetry [$SA = (R^{\uparrow \uparrow} - R^{\downarrow \downarrow})/(R^{\uparrow \uparrow} + R^{\downarrow \downarrow})$] for the Nb, Nb/Al and Nb/Au samples, measured at 42 mT, are shown in Fig. \ref{fig:PNR}, alongside theoretical fits. The PNR Bragg peak spacing ($\Delta Q = 2\pi / t $) shows that the superlattice structures are close to the nominal layer thicknesses, and the spin asymmetry serves to highlight the differences in scattering intensity between $R^{\uparrow \uparrow}$ and $R^{\downarrow \downarrow}$ induced by Meissner screening. For data taken at 20 K for the Nb/Al sample, we observe no observable spin asymmetry, as expected (supplemental Fig. S3 \cite{SM}). Specular and off-specular background XRR are collected in order to further validate the neutron reflectometry results as shown in supplementary Fig. S1 \cite{SM}. We also collect x-ray rocking curves to provide additional qualitative information about sample roughness, and the data can be seen in supplementary Fig. S2 \cite{SM}.

\begin{figure*}
    \centering
    \includegraphics[width=0.9\textwidth]{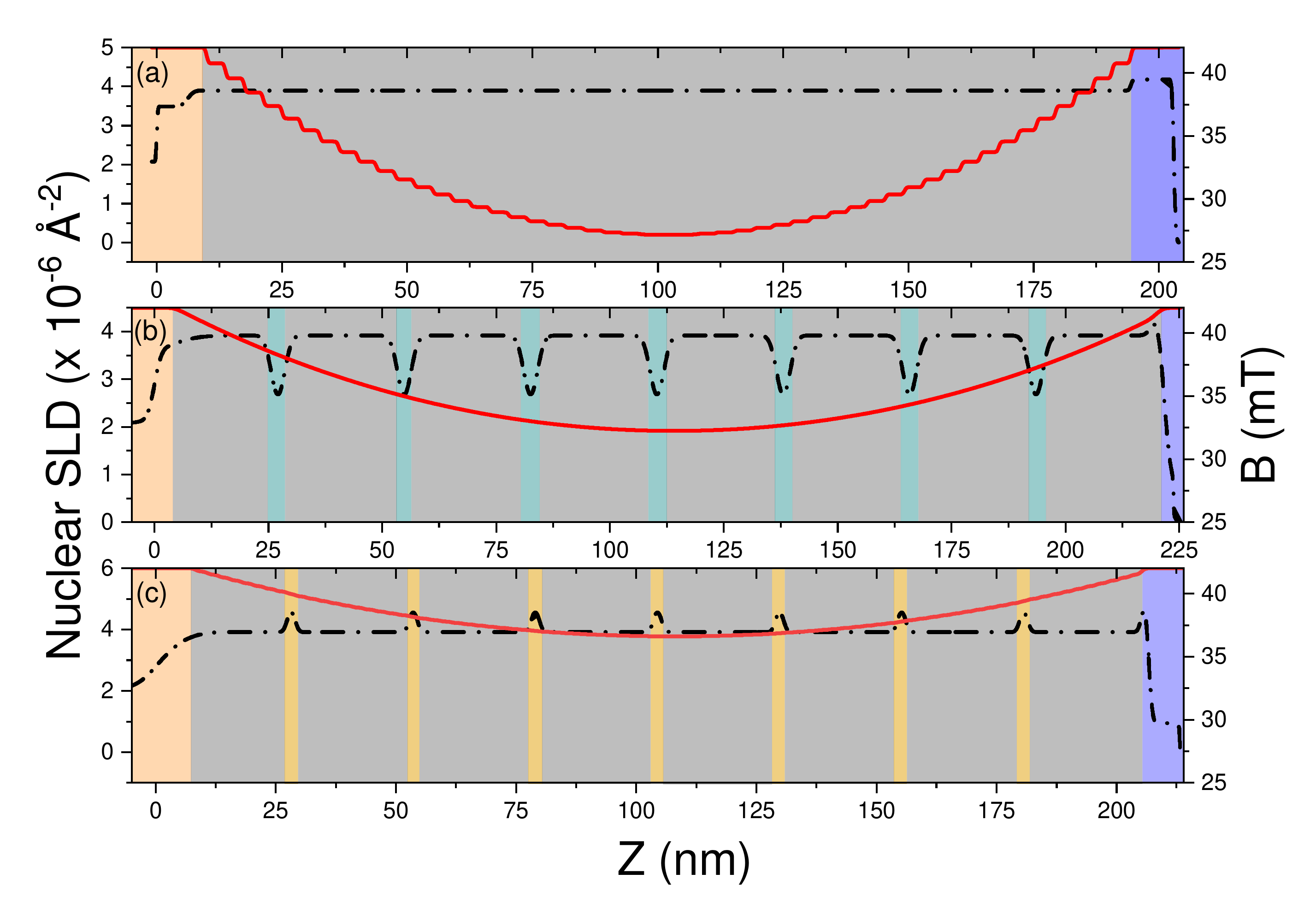}
    \caption{Nuclear SLD (black, left axis) and magnetic field (red, right axis) for (a) 200 nm Nb, (b) Nb/Al (grey/teal) and (c) Nb/Au (grey/gold) superlattices as a function of depth in the sample for B = 42 mT and T = 3 K. \textit{Z} = 0 refers to the Si substrate surface.  A thin native SiO\textsubscript{x} layer is included between the substrate and Nb (both Si and SiO\textsubscript{x} are denoted by the beige shaded region). An additional layer is needed to account for oxidation at the top Nb surface. A condensation of gas on the sample surface occurred at low temperatures for the Nb/Au sample. Both the Nb oxidation and condensation layers are denoted by the purple shaded region.}
    \label{fig:SLD}
\end{figure*}

The scattering length density (SLD) profiles that yield the best theoretical fits for the $R^{\uparrow \uparrow}$ and $R^{\downarrow \downarrow}$ data in Figs. \ref{fig:PNR} (a), (c), and (e) are shown in Fig. \ref{fig:SLD}. When fitting the superlattice structure, each elemental layer is constrained to the same nuclear SLD and thickness. The nuclear scattering length densities are in strong agreement with the bulk values for each layer. We model the magnetic field expulsion as a function of depth (where \textit{z} is distance from edge of the superconductor) in the superlattice as determined from the London equation \cite{PenetrationNIST},

\begin{equation} \label{eq:london}
    B(z) = B_{0} \cosh{\bigg( \frac{z}{\lambda_\text{L}}} -\frac{d_{s}}{2 \lambda_\text{L}} \bigg)  \cosh{\bigg(\frac{d_{s}}{2 \lambda_\text{L}}\bigg)}^{-1},
\end{equation}

\noindent where $d_s$ and $z$ are the thickness of the superconductor and distance from the surface, respectively. In this model, the external field ($B_0$) is fixed to the value measured with a Hall probe, while $\lambda_\text{L}$ is a fitting parameter. Our modeling assumes that there is no observable flux trapped by the formation of superconducting vortices which is reasonable given that the applied field is well below $H_{c1}$, and the fits without accounting for vortices are of excellent quality. In prior PNR reports vortices were not needed to explain the data for fields under 100 mT \cite{PNRNbAl}. Furthermore, based upon our previously measured Fraunhofer patterns in Josephson junctions with a Nb/Al superconducting electrode, we found no degradation of the pattern (which is well known to occur when flux is trapped) for fields as large as 120 mT \cite{PhysRevB.96.224515}, which is far larger than the 42 mT applied in this work. Our fits for the 200 nm single layer of Nb yield $\lambda_\text{L}$ $=$ 96.2 $\pm$ 9.2 nm (uncertainties on fit parameters correspond to 2 $\sigma$), which is in good agreement with prior experimental reports \cite{PenetrationPNR,PenetrationNIST,PNRNbCuNi}. We find $\lambda_\text{{L}}$ $=$ 145 $\pm$ 25 nm and 190 $\pm$ 26 nm for the Nb/Al and Nb/Au superlattices, respectively. Our measurement of $\lambda_\text{L}$ for the Nb/Al sample is consistent with results reported values of 180 nm by Han \textit{et al.}, despite their Nb thickness being significantly thinner \cite{PNRNbAl}. Furthermore, Madden \textit{et al} found that a penetration depth of approximately 185 nm was needed to model the inductance of their SQUID devices \cite{Madden_2018}. The increase in $\lambda_\text{L}$ for the superlattice samples is evident from a qualitative inspection of the spin asymmetry data near $Q = 0.14$ $\text{nm}^{-1}$, where the amplitude is significantly larger in Nb than in the superlattices; additionally the oscillations in the spin asymmetry are noticeably less clear at higher $Q$ in the superlattices compared to those for  pure Nb. We have also carried out additional model fitting on the superlattice samples where the penetration depth is fixed at several values, and all other parameters are again fit, in order to qualitatively demonstrate which features drive the determination of the penetration depth and associated error bar size (supplementary Fig. S4) \cite{SM}.

\begin{figure}
        \includegraphics[width=0.5\textwidth]{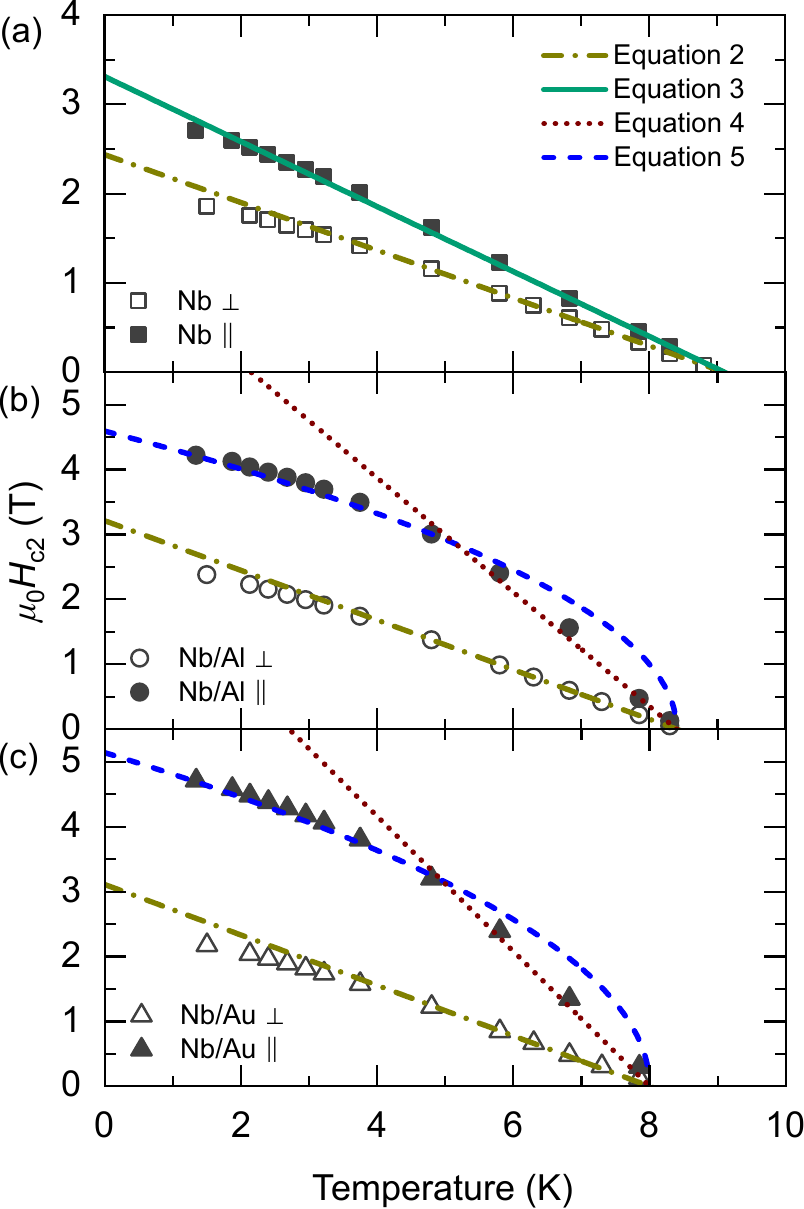}

        \caption{Temperature dependence of the upper critical field, $H_{c2}$, as determined from resistance as a function of out-of-plane ($\perp$) and in-plane ($\parallel$) magnetic field for (a) Nb, (b) Nb/Al and (c) Nb/Au samples. The data are modeled using the Ginzberg-Landau theory, where Equation \ref{eq:coherence3D} provides an estimate for the isotropic coherence length of the Nb in all samples from the $\perp$ magnetic field data. To model the $\parallel$ magnetic field data, Equations \ref{eq:Hc2OOP} and \ref{eq:Hc2IP} are fit over the entire temperature range to model 3D behavior in the Nb and 2D behavior in the superlattices, respectively. Equation \ref{eq:coherence3Dsuperlattice} is fit only to the high temperature $\parallel$ magnetic field data of the superlattices, where the temperature dependence is linear, to determine an estimate for the effective out-of-plane coherence length, see text. } 
        \label{fig:Hc2}
\end{figure}

\subsection{Electrical Transport}

To determine the effect of the thin Al and Au layers on the coherence lengths, we measure $H_{c2}$ as a function of temperature for the magnetic field applied both in- and out-of-plane, as shown in Fig. \ref{fig:Hc2}. Example resistance as a function of field data sets can be found in supplementary Fig. S5 \cite{SM}, which is then used to determine the $H_{c2}$ values for Fig. \ref{fig:Hc2}. The measured $T_{\text{critical}}$ at zero applied field for each sample, used in later calculations, is given for each sample in Table \ref{tab:SCparameters}.  





We use the $H_{c2}$ measurements to estimate the coherence lengths in the Nb and superlattices using the Ginzburg-Landau theory. In all our samples there exists the isotropic coherence length of Nb, $\xi_{\text{Nb} 0}$, however in the superlattices there is also an effective coherence length perpendicular to the layers, $\xi_{\perp 0}$. This additional effective coherence length arises in the superlattices as the currents perpendicular to the plane are strongly influenced by the coupling of the Nb layers through the normal-metal layers and by the impedance provided by the Nb-Al or Nb-Au interfaces. As previous work has shown, the anisotropy in superlattices can be treated in analogy to high temperature cuprate layered superconductors \cite{PhysRevLett.66.2826, tinkham_introduction_2004}.

We estimate the zero temperature in-plane coherence length ($\xi_{\parallel 0}$), from the out-of-plane field data in Fig \ref{fig:Hc2} and fitting to,

\begin{equation} \label{eq:coherence3D}
\mu_0H_{c2\perp}(T) = \left(\frac{\Phi_0}{2\pi \xi_{\parallel 0}^2}\right)\left( 1-T/T_{\text{critical}} \right),
\end{equation}

\noindent where $\Phi_0$ is the flux quantum. For the Nb film and superlattices, $\mu_0H_{c2\perp}(T)$ is linear just below $T_{\text{critical}}$ (where the Ginzburg-Landau theory applies) and 
the fitted coherence lengths are: $\xi_{\parallel 0} = 11.6 \pm 0.1$~nm, $10.1 \pm 0.1$~nm, and $10.3 \pm 0.1$~nm,  for Nb, Nb/Al, and Nb/Au, respectively.  For the Nb sample, we can identify $\xi_{\parallel 0} \equiv \xi_{\text{Nb} 0}$, where $\xi_{\text{Nb} 0}$ is the isotropic coherence length of Nb. In the superlattices, $\xi_{\parallel 0}$ is still consistent with $\xi_{\text{Nb} 0}$ in the systems, which has been slightly modified due to the \textit{N} layers. 

As expected for a 3D superconductor, the upper critical field for Nb in the parallel-field configuration can be described by,
\begin{equation}\label{eq:Hc2OOP}
\mu_0H_{c2\parallel}(T) \propto \left( 1-T/T_{\text{critical}} \right).
\end{equation}
$\mu_0H_{c2}$ is approximately 40\% larger at zero temperature for the parallel field configuration compared to the perpendicular field. This occurs since $\lambda_\text{L}$ is not much smaller than the sample thickness, thus at large enough fields, the supercurrents of each vortex in the film make contact with the film surfaces \cite{SuperlatticeCrossover}, which we note is unaccounted for by the Ginzburg-Landau theory.

For the superlattices in an in-plane field, we must consider two separate temperature regimes. Close to $T_{\text{critical}}$, $H_{c2\parallel}(T)$ is linear and consistent with Eqn. 3, but below $T_{\text{critical}}/2$ follows a square root dependence, as is highlighted by the fits in Fig. \ref{fig:Hc2} (b) and (c). The underlying high-field physics in these samples is again driven by the size, shape, and position of vortices in the system. 

For an in-plane field at $T$ close to $T_{\text{critical}}$, we observe behavior similar to that of a 3D superconductor since the diameter of the vortex cores is so large that the vortices can be viewed as averaging over the entire superlattice. And thus, the linear dependence of $H_{c2\parallel}(T)$ can then be fit to the Ginzburg-Landau expression,

\begin{equation} \label{eq:coherence3Dsuperlattice}
\mu_0H_{c2\parallel}(T) = \left(\frac{\Phi_0}{2\pi \xi_{\text{Nb} 0} \xi_{\perp 0}}\right)\left( 1-T/T_{\text{critical}} \right),
\end{equation}

\noindent where the proportionality constant can now be interpreted in terms of a phenomenological out-of-plane coherence length, $\xi_{\perp 0}$, associated with properties of the \textit{N} layers and niobium. We note that $\xi_{\perp 0}$ is not to be understood as an effective coherence length inside the \textit{N} layers. Using the $\xi_{\parallel 0}$  values determined for each superlattice (approximately equal to $\xi_{\text{Nb} 0}$), fits to Eqn. \ref{eq:coherence3Dsuperlattice} are shown in Fig. \ref{fig:Hc2} (b) and (c). We find that $\xi_{\perp 0} = 4.4 \pm 0.1$~nm, and $3.8 \pm 0.1$~nm for Nb/Al, and Nb/Au, respectively. 

At temperatures below $T_{\text{critical}}/2$, $H_{c2\parallel}(T)$ no longer follows the linear form, and instead is best described by the square root dependence expected for 2D superconductors,
\begin{equation}\label{eq:Hc2IP}
\mu_0H_{c2\parallel}(T) \propto \left( 1-T/T_{\text{critical}} \right)^{1/2}.
\end{equation}
\noindent In this regime, the cores of the vortices are confined to the \textit{N} layers, and do not significantly penetrate into the Nb layers. 

The Ginzburg-Landau theory predicts that close to $T_{\text{critical}}$,

\begin{equation} \label{eq:coherence2Dsuperlattice}
\mu_0H_{c2\parallel}(T) = \left(\frac{\sqrt{3} \Phi_0}{\pi d \xi_{\text{Nb}0}}\right) \left( 1-T/T_{\text{critical}} \right)^{1/2} ,
\end{equation}

where $d$ is the thickness of the Nb layers. Setting $d=25$~nm and using our previously determined $\xi_{\text{Nb} 0} = 10.1$~nm for the Nb/Al superlattice, Eqn. \ref{eq:coherence2Dsuperlattice} predicts $\mu_0H_{c2\parallel}(0 K) \approx 4.5$~T, in decent agreement with the measured values of $4.6 \pm 0.1$~T and $5.1 \pm 0.1$~T for Nb/Al and Nb/Au, respectively. This agreement between experiment and theory is further supportive of our interpretation that the superlattices behave as 2D superconductors.  

Finally, we summarize our experimental findings for each relevant superconducting parameter (e.g. $T_{\text{critical}}$, $\lambda_\text{L}$, and the discussed in- and out-of-plane upper critical fields and coherence lengths ) in Table \ref{tab:SCparameters}.

\begin{table*}
\caption{Summary of experimentally measured superconducting parameters for Nb, Nb/Al and Nb/Au samples. $T_\text{critical}$ is the critical temperature measured by electrical transport at zero applied field. $\lambda_\text{L}$ is the London penetration depth, as measured by PNR. $H_{c2}$ (0 K) is the upper critical field at zero temperature extracted from Figure \ref{fig:Hc2}. $\xi_{\parallel 0}$ is the experimental in-plane coherence length, which gives an estimate of $\xi_{\text{Nb} 0}$, the isotropic coherence length of Nb. $\xi_{\perp 0}$ is the phenomenological parameter describing the anisotropic coherence perpendicular to the plane in the superlattices.}

\begin{tabular}{c|c|c|c|c|c|c}
Sample & $T_\text{critical}$  & $\lambda_\text{L}$ (3 K)  & $\mu_0 H_{c2\parallel}$ (0 K)  & $\mu_0 H_{c2\perp}$ (0 K) & $\xi_{\parallel 0}$ ($\xi_{\text{Nb} 0}$) (0 K)  & $\xi_{\perp 0}$ (0 K)  \\ 
    & (K)    &  (nm)  & (T)    &  (T)    &  (nm)  & (nm)   \\ \hline
Nb   & 9.10 $\pm$ 0.05  & 96.2 $\pm$ 9.2      & 3.29 $\pm$ 0.02   & 2.32 $\pm$ 0.02   &  $11.6\pm0.1$  & --      \\ \hline
Nb/Al & 8.40 $\pm$ 0.05 & 145 $\pm$ 25      & 4.6 $\pm$ 0.1        & 3.02 $\pm$ 0.03    &$10.1\pm0.1$   & 4.4 $\pm$ 0.1      \\ \hline
Nb/Au & 8.00 $\pm$ 0.05 & 190 $\pm$ 26      & 5.1 $\pm$ 0.1           & 2.81 $\pm$ 0.04 &$10.3\pm0.1$   & 3.8 $\pm$ 0.1     
\end{tabular}

\label{tab:SCparameters}
\end{table*}

\section{Discussion}

From the combination of PNR, XRR, and electrical transport, we develop a consistent picture of the means by which the superconducting properties are affected by the structural differences between the \textit{S}/\textit{N} superlattices and uniform superconducting films of similar thickness.  The fits for the PNR and XRR show that the repeated layers in our superlattices are uniform with near nominal deposition thicknesses and SLD values close to bulk. The XRR fitted roughnesses of the topmost surface for the Nb/Al and Nb/Au superlattices (0.60 and 0.69 nm, respectively) are smaller than those obtained for the Nb film (1.95 nm). The improved surface roughness of the superlattices is qualitatively consistent with trends reported for similar Nb and Nb/Al samples with root-mean-squared roughnesses of 0.53 and 0.23 nm, respectively, obtained from atomic force microscopy over an area of 1~$\mu$m$^{-2}$ \cite{NbAlarea}. Note that the roughness values obtained from reflectivity measurements are typically larger than those obtained from atomic force microscopy since they represent an average of interdiffusion, local roughness, and large-scale features averaged across the sample plane. We also note that the Nb/Al sample characterized by Wang \textit{et al.} had a superlattice repeat of 3 \cite{NbAlarea}, and not 7 as we use in this work. We also observe satellite (Yoneda) peaks in the rocking curve for the Nb film at $\theta$ positions of $\theta_C$ (where $\theta_C$ corresponds to the critical angle for total internal reflection in Nb) and $2\theta$ - $\theta_C$.  Yoneda peaks, however, are not apparent in comparable rocking curves for the  Nb/Al and Nb/Au samples.  In general, Yoneda scattering results from a resonant enhancement of scattering from faceted surfaces. Since these features can be qualitatively linked to surface roughness, the pronounced Yoneda wings in rocking curves for the Nb film suggest that its surface is significantly rougher than the surfaces of the superlattices \cite{Sinha} (see supplementary Fig. S2 \cite{SM}).

Prior studies demonstrated that the thin intermediary layers (Al and Au in this work) disrupt the columnar growth of Nb, resulting in reduced surface roughness in Nb/\textit{N} superlattices \cite{NbAlplanarization, NbSiplanarization}. This conclusion is further supported by XRR measurements of the off-specular background for the superlattices and film. While the off-specular scattering for the Nb film is mostly featureless (supplemental Fig. S1c \cite{SM}), the scattering for the superlattices has finite-size oscillations (with constant spacing in $Q$) and diffraction peaks that mirror the features present in the specular reflectivity (supplemental Fig. S1 \cite{SM}).  The presence of these off-specular oscillations suggests that the in-plane interface roughness originates in the layers near the substrate and is replicated from one layer to the next (i.e., conformal) \cite{Sinha,ConfromalSiMo,ConfromalZabel}. The superlattice interfaces are presumably well-defined on a local scale, and the resulting smooth surface contrasts with that of the Nb film. 

Experimental measurements of the penetration depth ($\lambda_\text{L}^{\text{experimental}}$) are known to be larger than the intrinsic London penetration depth ($\lambda_\text{L}^{\text{intrinsic}}$), due to impurity defects. A relationship between the intrinsic London penetration depth and that measured by experiment has been derived by Pippard \cite{Pippard} and demonstrated by Zhang \textit{et al} \cite{PenetrationNIST}, as

\begin{equation}\label{eq:Intrinsic}
\lambda_{\text{L}}^{\text{experimental}} = \lambda_{\text{L}}^{\text{intrinsic}}\sqrt{1 + \frac{\xi^{\text{intrinsic}}_{\text{Nb}}}{l}},
\end{equation}

\noindent where \textit{l} is the electron's mean free path (5.5 nm based upon previous resistivity measurements \cite{Bass_2007,PenetrationNIST}), and  valid when $l << \xi$. Using the experimental value for $\lambda_{\text{L}}$ and $\xi^{\text{intrinsic}}_\text{Nb} = 41$~nm as reported by Weber \textit{et al} \cite{PhysRevB.44.7585}, we calculate $\lambda_{\text{L}}^{\text{intrinsic}}$ = 33.1 nm. This value is reasonably consistent with the reported value of Zhang \textit{et al} \cite{PenetrationNIST}.

Our measured penetration depth for Nb/Al is in agreement with prior reports by Han \textit{et al.} \cite{PNRNbAl} who employ PNR on Nb/Al superlattices and found $\lambda_L$ = 180 $\pm$ 20 nm, though in this case the Al normal-metal layer was even thinner than in our work. 
Furthermore, Madden \textit{et al} \cite{Madden_2018} could simulate the inductance of their SQUIDs only if they used a longer $\lambda_\text{L}$ (185 nm) for Nb/Al (with 3 multilayer repeats), which supports our findings with an entirely independent technique. Finally, during the preparation of this work, it has been reported that a larger penetration depth of approximately 190 nm is necessary to satisfactorily model the flux through Josephson junctions where a Nb/Au superlattice serves as one of the electrodes \cite{doi:10.1063/1.5140095,satchell2020ptcobpt}. While this much longer penetration depth in \textit{S}/\textit{N} superlattices is perhaps surprising, our measurements are in strong agreement with multiple independent groups and measurement methods.

\section{Conclusions}

In summary, we have studied the effect of adding thin normal-metal layers between Nb layers by directly measuring the London penetration depth and Ginzburg-Landau coherence length. 
We find that the superlattices have a significantly longer penetration depth compared to Nb films of similar thickness. To further characterize the superlattices, we determine an effective coherence length for currents perpendicular to the plane, which we find to be less than half that of the intrinsic coherence length of Nb. The result of this is a decoupling of the \textit{S} layers in the superlattices, such that below $T_{\text{critical}}/2$ the superlattices act as layered 2D superconductors, unlike the single Nb film that displays 3D behavior across all temperatures measured.  
The changes 
suggest that the addition of thin \textit{N} layers between the \textit{S} layers weakens the superconducting coherence in the out-of-plane direction despite the expectation that the \textit{N} layers would be fully proximitized given that the coherence length ($\xi_{\text{Nb}0}$) is large with respect to the thickness of the N layers. This weakening of the out-of-plane coherence causes the superlattices to act as an anisotropic 2D superconductor, unlike the Nb sample which displays isotropic behavior. 
The lower surface roughness associated with using a \textit{S}/\textit{N} superlattice compared to a single \textit{S} layer is valuable in Josephson junction applications given the overall improved quality of the junction, however, this has the unintended consequence of significantly changing characteristic superconducting properties. To properly model such systems, one must account for changes in $\lambda_\text{L}$, $\xi_\text{GL}$, and $T_\text{critical}$-- which we independently and directly measure in this work.

The data associated with this paper are openly available from the NCNR and University of Leeds data repositories \cite{Data}.

\begin{acknowledgments}

We thank Alexander Grutter and Yury Khaydukov for helpful discussions, Josh Willard for initial characterization of the superlattices, and Tanya Dax, Joseph Dura and Brian Maranville for assistance with the PNR measurements. We are grateful for the extensive constructive feedback received during the review process that resulted in improved clarity of this manuscript. P.Q. acknowledges support from the National Research Council Research Associateship Program. This project has received funding from the European Unions Horizon 2020 research and innovation programme under the Marie Sk\l{}odowska-Curie Grant Agreement No. 743791 (SUPERSPIN).

P. Quarterman and N. Satchell contributed equally to this work.

\end{acknowledgments}

\bibliography{SCsuperlattices}
\end{document}


\begin{center}
\large{Supplementary Material for ``Distortions to the penetration depth and coherence length of superconductor/normal-metal superlattices'' by P.~Quarterman \textit{et al}.}%
\end{center}


X-ray reflectivity is collected on the samples uing a Cu-K$\alpha$ source (supplementary Fig. \ref{fig:XRR}(a)), and the resulting fits to the data yield scattering length density profiles, shown in supplementary Fig. \ref{fig:XRR}(b), in strong agreement with the polarized neutron reflectometry results. From the PNR and XRR we find that thicknesses for  Nb/Al and Nb/Au superlattice samples are 25.2 nm / 2.5 nm and 23.6 nm / 1.7 nm, respectively. The single layer of Nb is found to be 188 nm. Off-specular scattering is collected to use for background subtraction by offsetting $\theta$ and scanning $\theta - 2\theta$ in order to sample the $Q_z$ dependence of the in-plane disorder (Fig. \ref{fig:XRR}(c)). For the off-specular background scattering in the Nb sample, we observe no $Q$-dependent oscillations, as expected. However, in the superlattices the background scans show strong periodic oscillations. These off-specular oscillations have been shown to be indicative of conformal roughness [1,2]. 

Rocking curves are shown for the Nb, Nb/Al, and Nb/Au samples in Fig. \ref{fig:XRR2}. The Nb rocking curve at 2$\theta = 1.2^\circ$  shows pronounced peaks (Yoneda wings) near $\theta \approx \theta_C$ and $2\theta$ - $\theta_C$, where $\theta_C$ is the critical angle. These Yoneda peaks are suggestive of significant surface roughness [3]. In contrast, the only pronounced peaks observed in the Nb/Al and Nb/Au rocking curves are the specular reflections, which are on top of broad diffuse scattering for the latter sample. In the Nb/Au rocking curve at 2$\theta$ = 1.3$\degree$, there are peak-like artifacts symmetric about the specular reflection. In the rocking curve measured at 2$\theta$ = 2.2$\degree$, however, these features do not shift outward in a manner consistent with Yoneda scattering. In qualitative  comparison to those of the Nb film, the surfaces of the Nb/N superlattices appear to be relatively smooth on a local scale. 

The spin asymmetry for the Nb/Al superlattice taken at 20 K is shown in supplemental Fig. \ref{fig:SA20K}, and as expected we see no observable magnetic contribution to the PNR. Similar observations are seen for the Nb and Nb/Au samples taken when measured above the superconducting critical temperature of the samples. We utilize a Markov chain Monte Carlo method, known as DREAM [4], which allows us to precisely determine uncertainties for numerous correlated parameters. We note that the uncertainties reported for our fit parameters do not account for deficiencies in the model or all of the systematic error that might be associated with the experiment. However, based on the theoretical plausibility of our model, and the overall goodness of fit we conclude that such unaccounted contributions to the total uncertainty are quite small, and that our model is a reasonable representation of reality. To demonstrate the robustness of our model-fitting of the penetration depth, we have carried out additional fitting where $\lambda_{L}$ is fixed to a series of values, and the scattering length densities, layer thicknesses, and interfacial roughnesses are fit (as used for main text). These alternative fits are shown in supplementary Fig. \ref{fig:AltModels}. In the Nb/Al sample, we find that the magnetic contribution to the model-fitting is dominated by the features near $Q = 0.14$ and 0.25 $\text{nm}^{-1}$, and when $\lambda_{L}$ is too small the theoretical model over estimates the spin asymmetry near $Q$ = 0.25 $\text{nm}^{-1}$, whereas if $\lambda_{L}$ is too large, the model does not capture the dip near $Q$ = 0.14 $\text{nm}^{-1}$. Thus, we can qualitatively understand the fit as balancing these two features for a minimization in $\chi^{2}$. We observe a similar balancing of the first low $Q$ dip (0.14 $\text{nm}^{-1}$) with oscillations at higher $Q$, in the spin asymmetry, for the Nb/Au sample.

We determine the upper critical fields ($H_{c2\parallel}$ and $H_{c2\perp}$) by measuring resistance as a function of field ($R$ vs. $H$); typical $R$ vs. $H$ curves for the field applied in-plane and out-of-plane are shown in supplemental Fig. 5. $H_{c2}$ is determined by taking the peak of the first derivative of the resistance with respect to field, and averaging the result obtained at the positive and negative field regime.

In the main text, we define several coherence lengths in our system. One such length that we do not consider in the main text is the normal state effective coherence length inside the \textit{N} layers ($\xi_N$). To provide an estimate for $\xi_N$ in the diffusive (dirty) limit for Al and Au, we can follow Buzdin [5],

\begin{equation} \label{eq:coherencenormal}
   \xi_N = \sqrt{\frac{\hbar D}{2 \pi k_B T}},
\end{equation}

\noindent where $D$ is the diffusion coefficient, $D= \frac{1}{3} v_\text{F} l$, and $v_\text{F}$ and $ l$ are the Fermi velocity and mean free path respectively. Values for $v_\text{F}$ and $ l$ are taken from Gall [6], with the limitation that $l$ is calculated by Gall at room temperature. We estimate the normal metal effective coherence length for Al and Au at 3~K, associated with the proximity effect, in the diffusive limit to be $\xi_N = 65$~nm and $\xi_N = 85$~nm.\\

\noindent[1]  H. Zabel, Appl. Phys. A. \textbf{58}, 159 (1994).

\noindent[2]  J. M. Freitag and B. M. Clemens, J. Appl. Phys. \textbf{89}, 1101 (2001).

\noindent[3]  D. E. Savage,  J. Kleiner,  N. Schimke,  Y. Phang,  T. Jankowski,  J. Jacobs,  R. Kariotis,   and M. G. Lagally, J. Appl. Phys \textbf{69}, 1411 (1991).

\noindent[4]  J. A. Vrugt, C. J. F. ter Braak, C. G. H. Diks, B. A. Robinson, J. M. Hyman and D. Higdon, Int. J. Nonlin. Sci. Num. \textbf{10} 273 (2009)

\noindent[5]  A. I. Buzdin, Rev. Mod. Phys. \textbf{77}, 935 (2005).

\noindent[6]  D. Gall, J. Appl. Phys \textbf{119}, 085101 (2016).

\begin{figure}[!p]
        \includegraphics[width=0.9\textwidth]{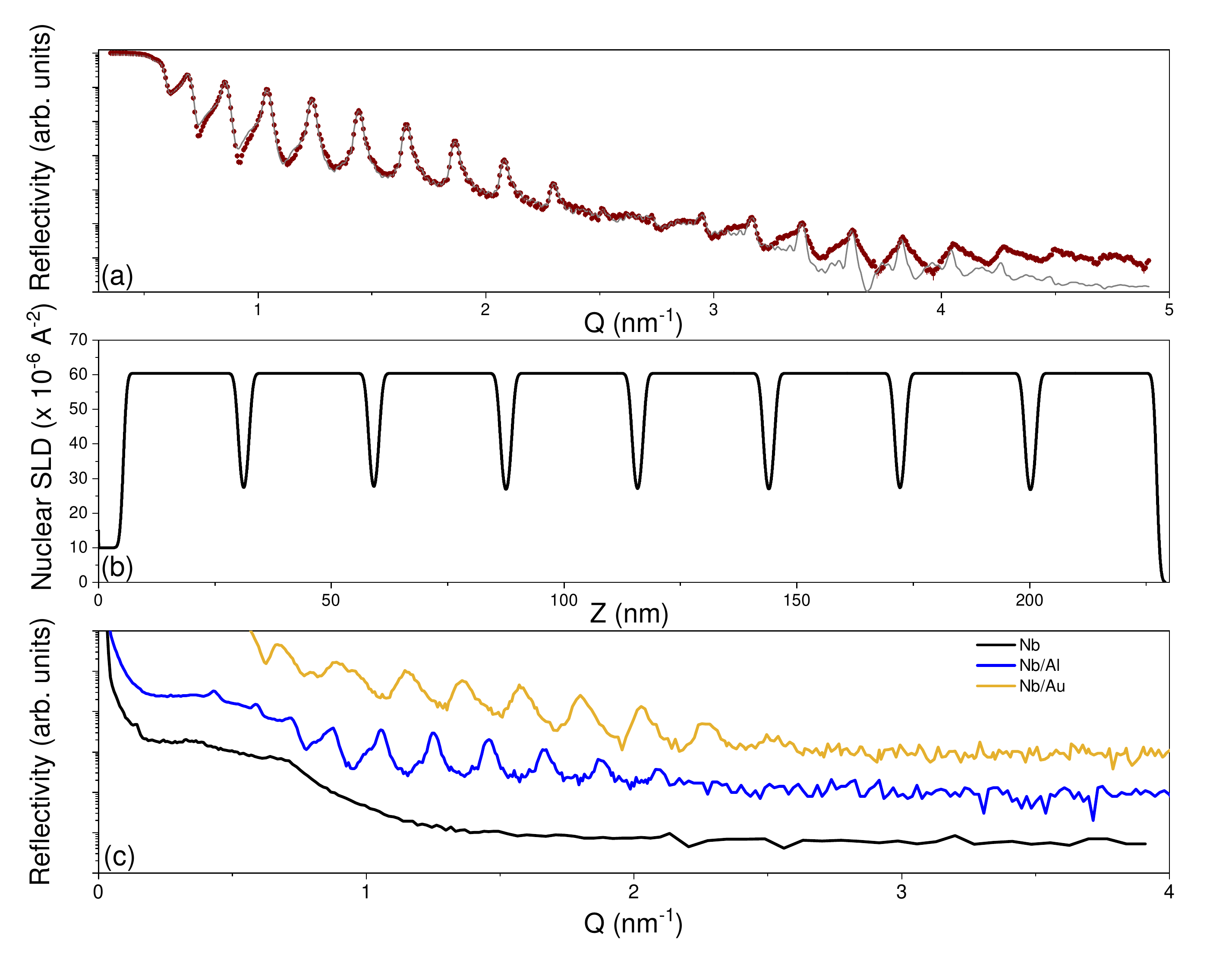}

        \caption{(a) XRR data and model for Nb/Al, with a corresponding SLD profile shown in (b). (c) Off-specular x-ray reflectivity for the Nb, Nb/Al and Nb/Au samples. The superlattices show pronounced oscillations in the off-specular signal, which are indicative of conformal roughness.}
        \label{fig:XRR}
\end{figure}

\begin{figure}[!p]
        \includegraphics[width=0.9\textwidth]{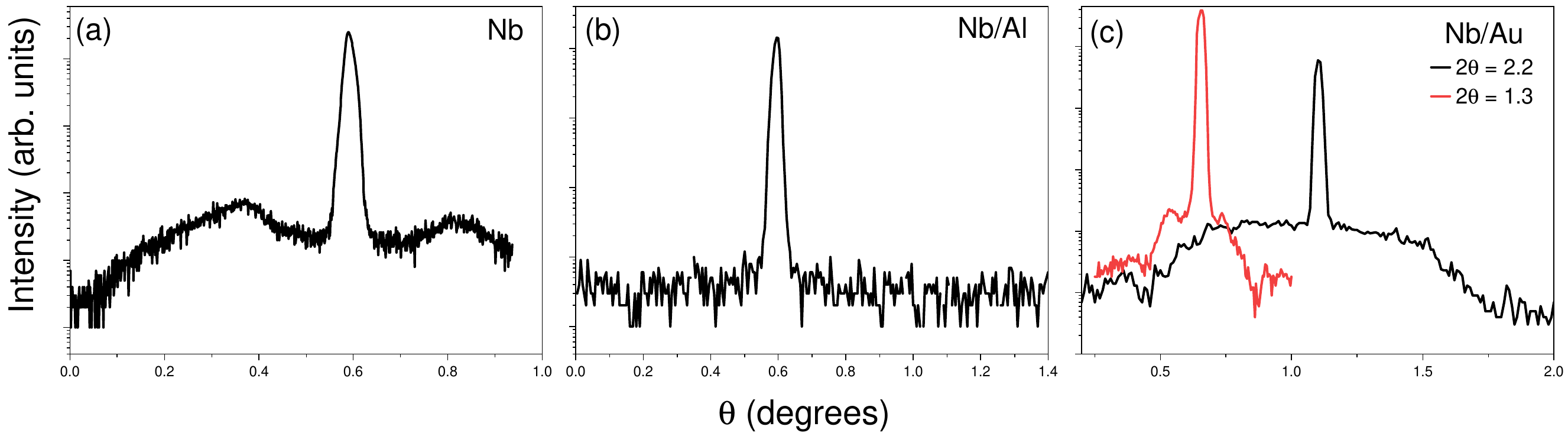}

        \caption{X-ray rocking curves of (a) Nb, (b) Nb/Al, and (c) Nb/Au samples. The Nb and Nb/Al rocking curves were obtained at 2$\theta =$ 1.2$\degree$. Rocking curves for Nb/Au were measured at 2$\theta =$ 1.3$\degree$ for direct comparison to (a) and (b) and at 2$\theta =$ 2.2$\degree$ at the maximum of a specular superlattice peak. A comparison of these two curves suggests that the apparent features in the diffuse scattering are not consistent with Yoneda scattering. }
        \label{fig:XRR2}
\end{figure}

\begin{figure}[!p]
    \centering
    \includegraphics[width=0.9\textwidth]{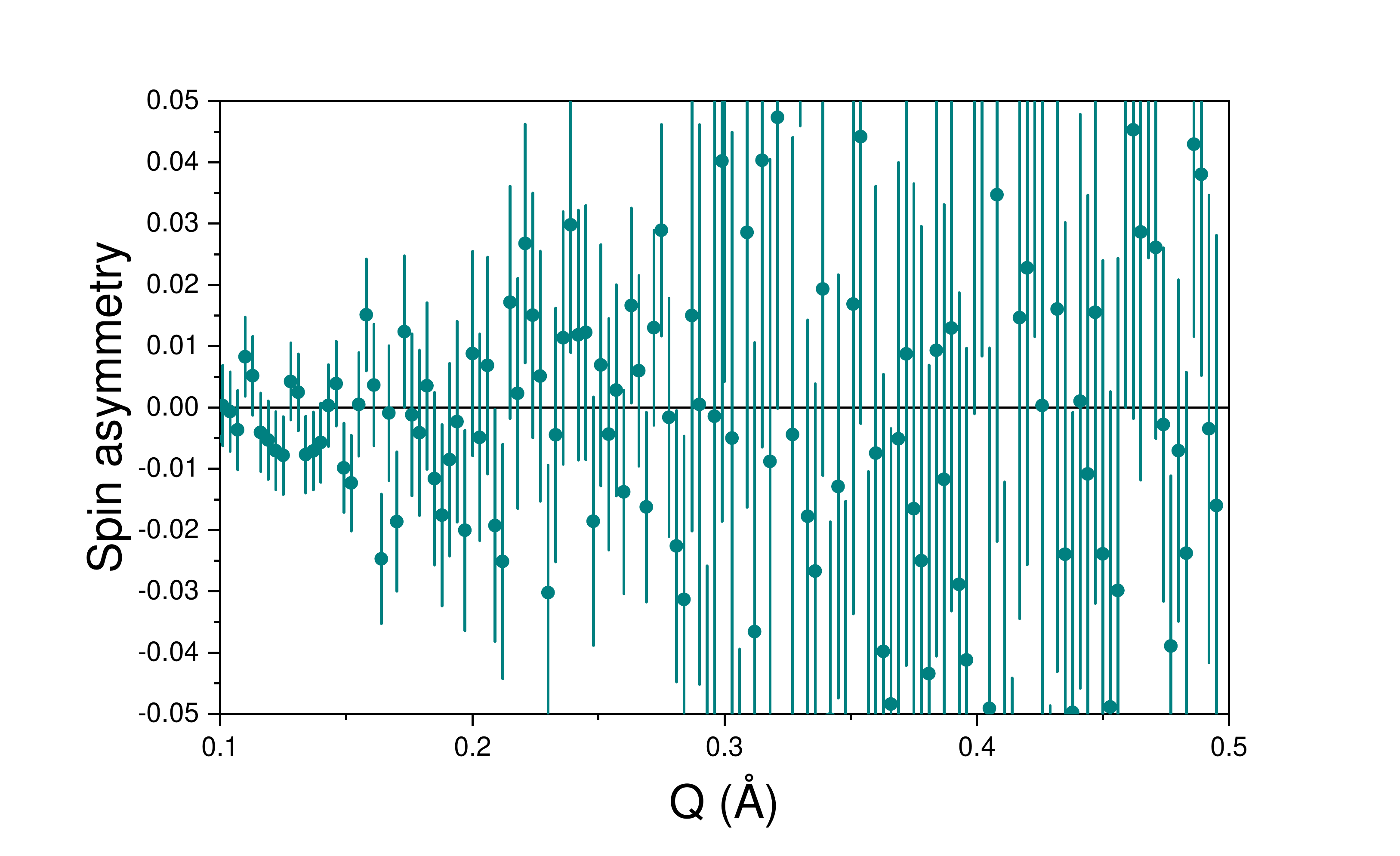}
    \caption{Spin asymmetry for Nb/Al superlattice collected at 20 K, which is above the superconducting critical temperature of the sample. Error bars are representative of 1$\sigma$}
    \label{fig:SA20K}
\end{figure}

\begin{figure}[!p]
    \centering
    \includegraphics[width=0.9\textwidth]{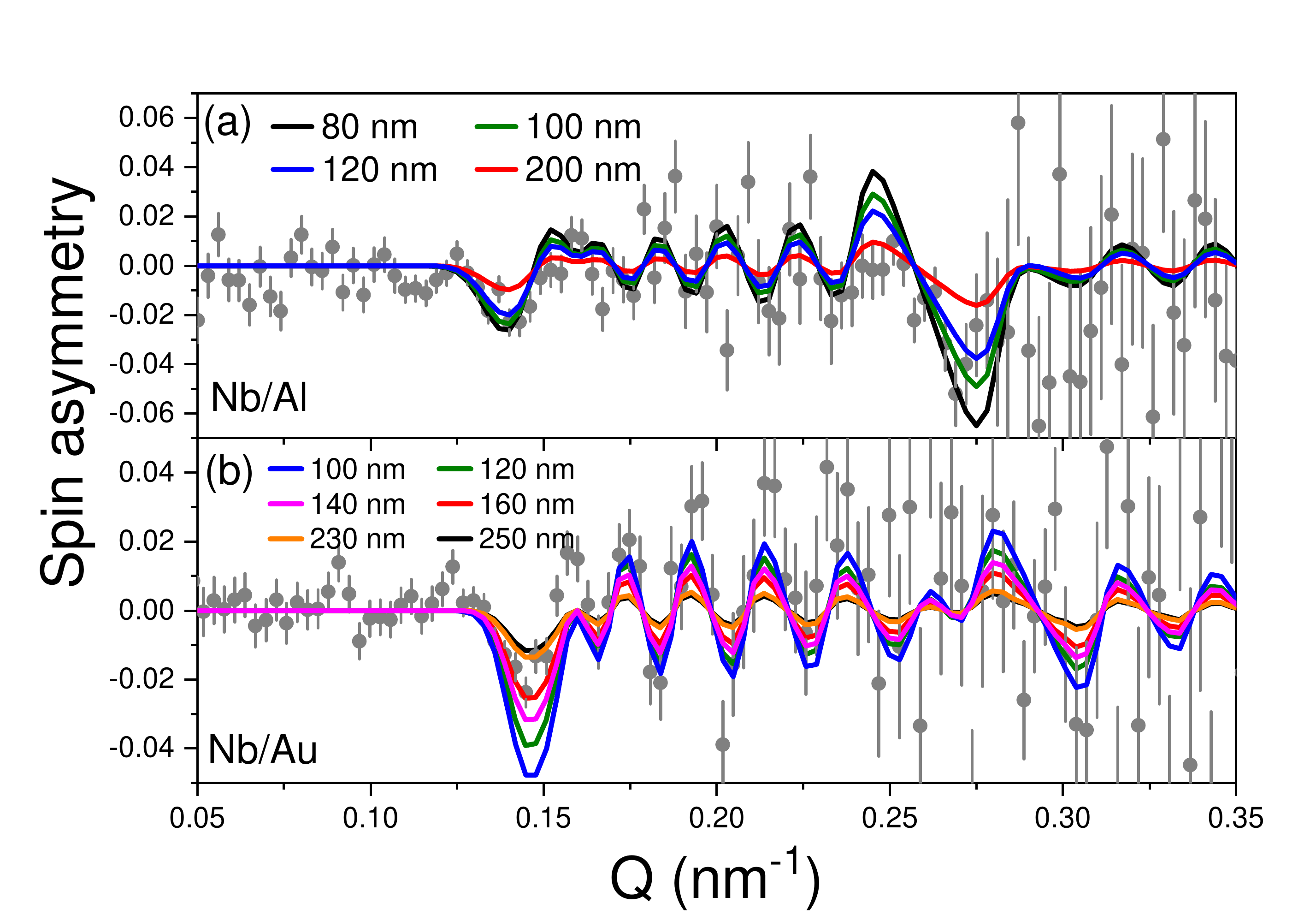}
    \caption{Spin asymmetry for (a) Nb/Al and (b) Nb/Au superlattice collected at 3 K, where modeling was done by fixing $\lambda_{L}$ (see legend) outside of the fit error range to highlight where model-fitting fails when $\lambda_{L}$ is either too large or too small. Error bars are representative of 1$\sigma$}
    \label{fig:AltModels}
\end{figure}

\begin{figure}
    \centering
    \includegraphics[width=0.9\textwidth]{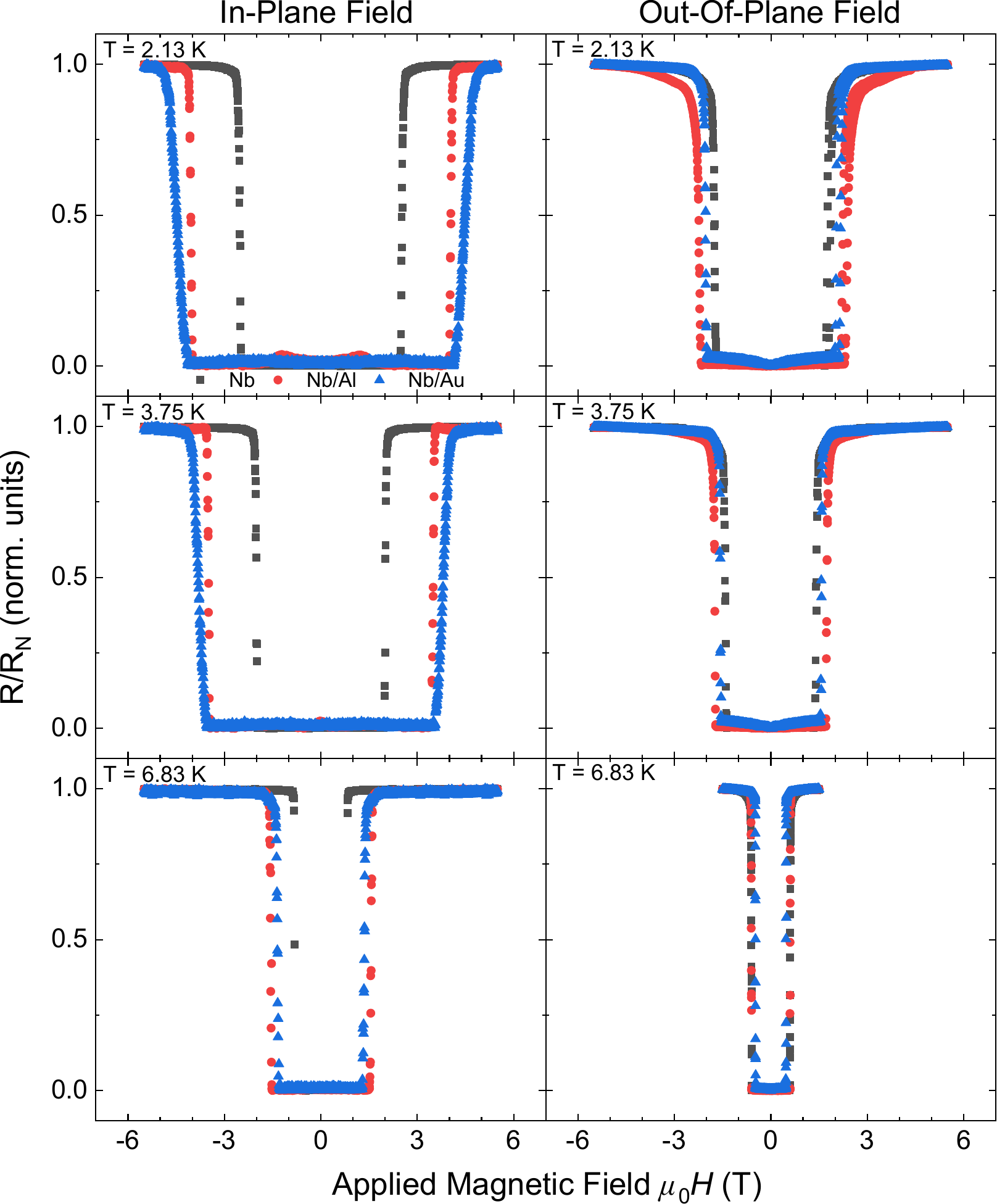}
    \caption{Resistance vs magnetic field for the Nb (black, squares), Nb/Al (red, circles), and Nb/Au (blue, triangles) samples to determine $H_{c2\parallel}$ and $H_{c2\perp}$.}
    \label{fig:my_label}
\end{figure}

\bibliography{SCsuperlattices2}